# PhysPort use and growth:
# Supporting physics teaching with research-based resources since 2011

Sarah B. McKagan[1], Linda E. Strubbe[2], Lyle J. Barbato[1], Adrian M. Madsen[1], Eleanor C. Sayre[2], Bruce A. Mason[1]
1. American Association of Physics Teachers
2. Kansas State University

## *Introduction*

### What is PhysPort?

PhysPort (www.physport.org) has become the go-to place for physics educators to learn to apply research-based teaching and assessment in their classrooms. Usage (defined as the total number of visits) has doubled every two years since the site was released in 2011, and 20% of all U.S. physics faculty and 7% of U.S. high school physics teachers are now verified educators on PhysPort. The lead author conceived of PhysPort in 2007 after meeting many physics instructors interested in incorporating results of physics education research (PER) in their classrooms but with no idea where to start. At the time, most PER results were only available in research journals, which required instructors' time and effort to find articles and read them, and there was no central place to learn about results and implications for classroom practice. Throughout its development, PhysPort has been based on user research: PhysPort staff interview physics instructors about their needs, design the site based on those needs, and conduct usability testing to see how they meet those needs. PhysPort is a joint product of the American Association of Physics Teachers (AAPT) and Kansas State University (K-State), with contributions from many universities, funded by the National Science Foundation (NSF), and a library within the ComPADRE.org digital library. The site (originally called the PER User's Guide) was first funded by an NSF grant in 2009 and released in November 2011. This article presents an overview of resources on PhysPort, discussion of research and development of the site, and data on the continuing growth of site usage.

### What can I find on PhysPort?

PhysPort contains many resources for physics educators. It is organized around "Expert Recommendations", "Teaching", "Assessments", "Workshops", and the "PhysPort Data Explorer" (see Figure 1 and Table 1). The Expert Recommendations section contains short articles answering the most common questions physics instructors have about teaching and assessment. Example topics include "What makes research-base teaching methods in physics work?" and "What can I do if students don't speak up in discussions with peers or the with the whole class?" The Teaching section contains guides to over 50 research-based teaching methods in physics as well as several collections of free open-source curricular materials for teaching physics. We define teaching methods broadly to include strategies, curricula, tools, and structures for courses. Overviews of over 50 research-based teaching methods help physics educators learn about these methods including when and how to use them, the research behind them and find example materials and classroom video of the methods being used. Educators can also find extensive downloadable implementation guides for some of the teaching method that include all relevant information for using the teaching method and links to resources for learning more. In most cases the teaching materials that accompany the teaching methods described on PhysPort are free, though some are available for purchase.

The Assessments section of PhysPort offers access to over 80 research-based assessments and includes all the necessary information to use the assessment such as the course level, format, duration, how to administer and score, and typical results. Educators can filter by a variety of categories to easily find the assessment they are looking for. PhysPort has assessments for most physics content topics, as well as assessments for non-content topics like beliefs and attitudes, self-efficacy, problem solving, and lab skills. The PhysPort Data Explorer enables educators to automatically analyze the results of many of the research-based assessments in physics and astronomy. This online tool quickly, easily and securely uploads a spreadsheet of student assessment results supplied by the instructor, and then visualizes the pre- and post-test results several ways, including the overall average normalized gain and effect size, a breakdown by topic, and a comparison to the national average for that assessment. The Data Explorer is free to use.

The "Workshops" tab includes two sets of video lessons. The Virtual New Faculty Workshop, a set of videotaped sessions from the live Workshop for New Physics and Astronomy Faculty. These sessions feature leaders in physics education presenting about research-based teaching techniques, curriculum, and pedagogy. The "Workshops" tab also includes the Periscope collection of video-based lessons for teaching assistants (TA) and learning assistant (LA) training. In these lessons, LAs and TAs watch and discuss videos of best practices in physics classrooms, practice interpreting student behavior and apply lessons learned to actual teaching situations to become more effective teachers. PhysPort also hosts a small collection of podcasts, "Learning about teaching physics", where each episode investigates a piece of research literature and how it can relate to your physics classroom.

Figure 1 and Table I describe the content available on PhysPort.

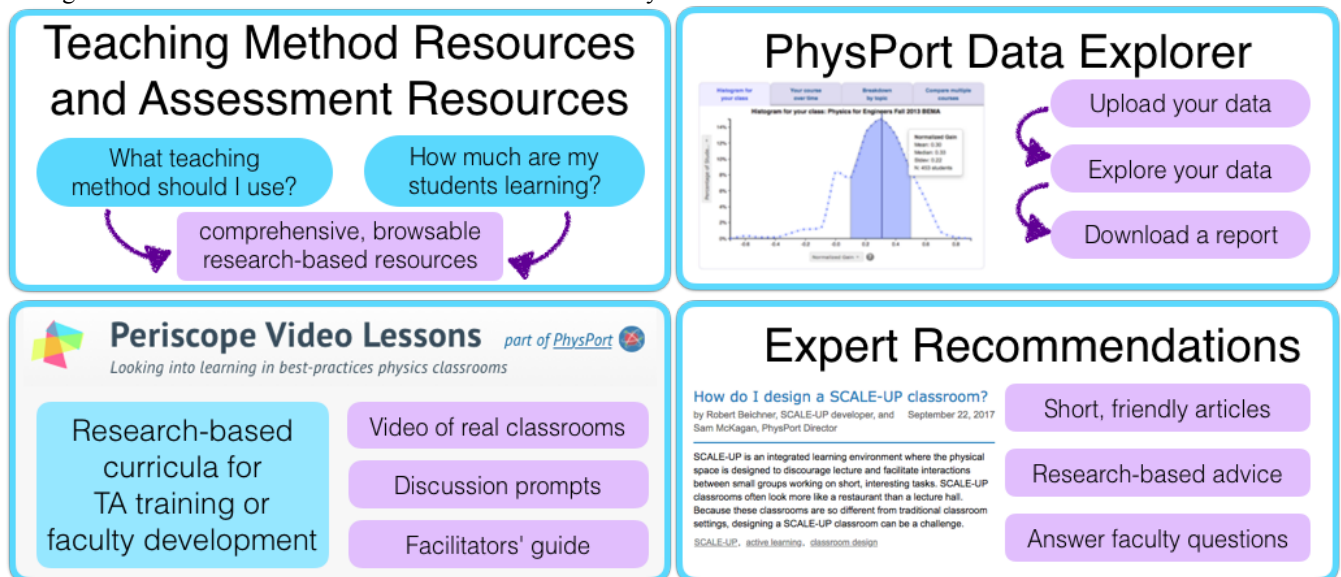

Fig. 1. Content available on PhysPort.

## *Research, design and development of PhysPort*

PhysPort's mission is supported by three kinds of research: (1) Fundamental research on physics instructors needs around teaching and assessment[1 2 3]; (2) Applied research on user experience with the PhysPort website[4]; and (3) Synthesis and fundamental research on effective practices in teaching, learning and assessment[5 6 7 8 9 10]. This research has been primarily among college / university physics instructors in the U.S. and focused on undergraduate teaching; we discuss other groups of PhysPort users below. Additionally, PhysPort enables research into student learning by creating tools that support developers of assessment instruments and curricula in their research efforts and helping them reach a broad audience. Development of PhysPort is ongoing, as the project team continually strives to improve the site to make it most useful to current and potential users. PhysPort developers are currently being funded to conduct research on how physics instructors use resources to support their teaching and how PhysPort impacts instructors' teaching practices.

## *Usage of PhysPort*

### **How well used is PhysPort?**

PhysPort usage is substantial and has increased annually since its release at the end of 2011. Figure 2 shows how key usage measures have increased from 2012 to 2018, determined from Google Analytics. During 2018, PhysPort had 84,000 visits[11], 56,000 unique visitors, and 1,500 users who visited the site five or more times. All of these measures have been increasingly annually since PhysPort's release.

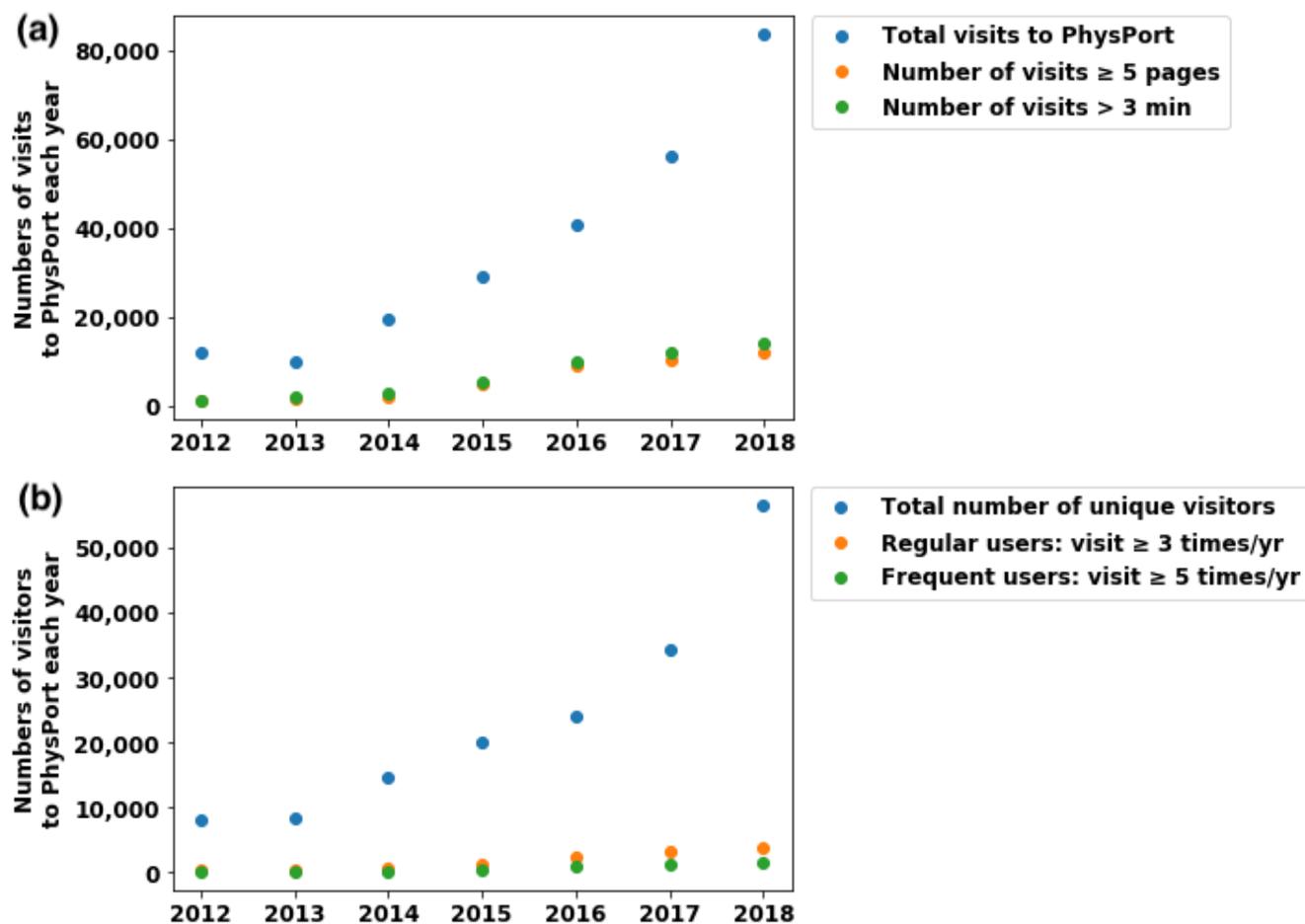

Figure 2. PhysPort usage over time, 2012-2018. (a) Number of visits to PhysPort each year. A visit to PhysPort ("session") may include looking at multiple pages within the site. We show total number of visits, number of visits where a user visited five or more pages, and number where the visit lasted longer than three minutes. (b) Number of visitors to PhysPort each year. A unique visitor is a user traceable to a unique IP address, who might visit the site multiple times. "Regular users" visit PhysPort more than three times in a year; "frequent users" visit more than five times in a year. Data is from Google Analytics.

## Which parts of PhysPort are most used?

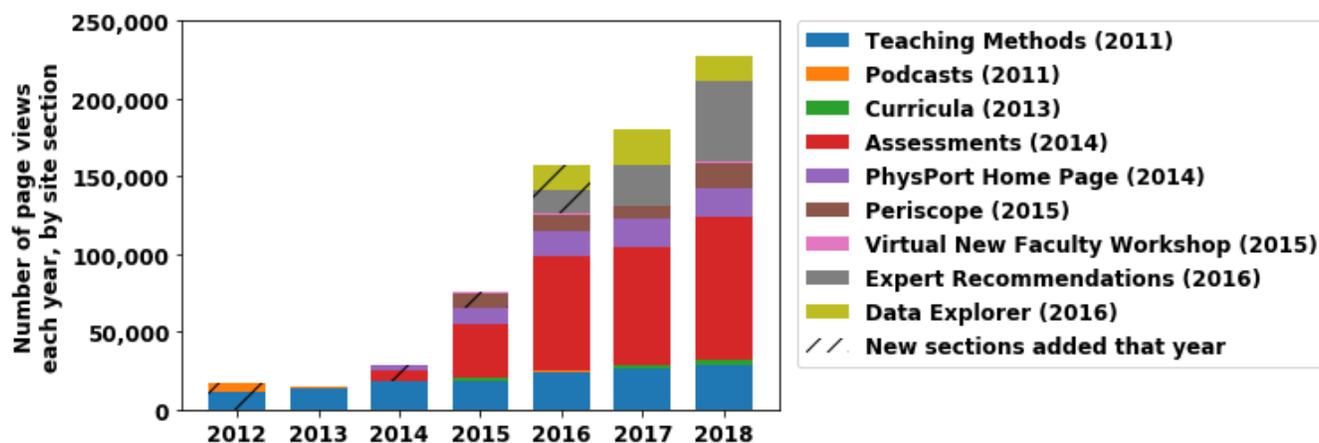

Figure 3. Total number of page views each year by site section. The dates in the legend indicate when the section was added to the site. Hatched regions indicate new sections added to the site that year. The Assessments and Teaching Methods sections receive the most total page views over time on PhysPort. Data is from Google Analytics.

Figure 3 shows the total number of page views for each section of the site. The Assessments section, where users can learn about and download research-based assessments, received the most page views on PhysPort in 2018. Of the 97 assessments published on PhysPort, 65 had more than 100 unique page views[12] and nine had more than 1,000 unique page views in 2018. The most popular assessment, the Force Concept Inventory (FCI), had 11,000 unique page views and was downloaded 1,400 times in 2018. Over the lifetime of PhysPort[13], users have downloaded almost 36,000 assessments.

The site section that has the second highest fraction of page views for 2018 is Expert Recommendations. For the 10 recommendations with over 1,000 unique page views in 2018, the average time spent on the page (for users who later visit another page on PhysPort) is five minutes. PhysPort's most popular recommendation in 2018, "Effect size: What is it and when and how should I use it?"[14], had 9,300 unique page views and an average time on page of nine minutes. These times are substantial, since web users usually click between pages very quickly: 80% of users leave a typical page within 70 seconds — and users tend to judge pages related to education and science particularly quickly[15]. We infer that these recommendations are capturing the attention of PhysPort visitors.

## Who is using PhysPort? How well is PhysPort reaching its target audiences?

The target audiences of PhysPort are two overlapping groups: physics/astronomy instructors at U.S. colleges / universities[16] and members of AAPT. The choice to prioritize U.S.-based physics/astronomy instructors and AAPT members is largely driven by the priorities of our funders, availability of research about these groups and their students, and our ease of access to these groups for needs and usability research. PhysPort has two other major populations of users as well: U.S. high school teachers and educators outside the U.S. We are pleased that these latter groups find PhysPort useful; however, we note that our user-centered research and design process has so far focused almost exclusively on U.S. college faculty—whose needs and contexts may be quite different. If funding were available in the future, we would be delighted to design PhysPort around the needs of U.S. high school teachers and educators outside the U.S. also.

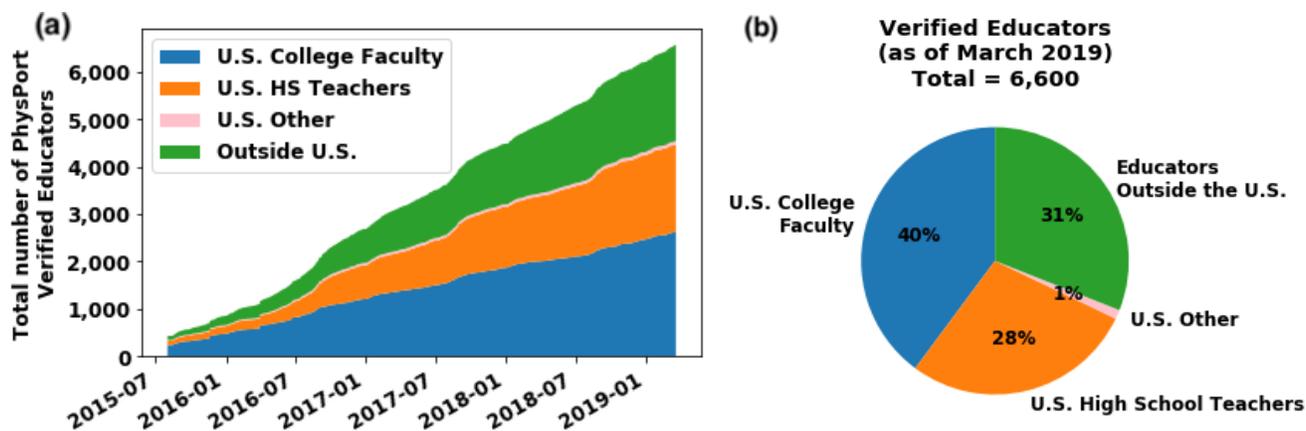

Figure 4. Verified Educators on PhysPort, broken down by population. (a) Growth of Verified Educators over time. (b) Verified Educators as of March 2019.

We primarily learn about who is using PhysPort by requesting demographic information from users who become "Verified Educators" so they can access restricted resources (download assessments, use the Data Explorer, and access Periscope workshops). Currently there are 6,600 Verified Educators, and the number has been growing steadily; see Figure 4. About one-third of Verified Educators are based outside the U.S., slightly more than one-third are at U.S. colleges, and slightly less than one-third are U.S. high school teachers. One-fifth of Verified Educators are AAPT members (see below for details). Only about 10% of Verified Educators are physics education researchers[17], indicating that PhysPort use extends well beyond the PER community.

In comparison to statistics from American Institute of Physics Statistical Research Center (AIP Statistics) (Table II), PhysPort is reaching a substantial fraction of our (overlapping) target audiences; we are also reaching substantial numbers of U.S. high school teachers and educators outside the U.S.

*1. U.S. college physics faculty*

U.S. college physics faculty are the main target audience of PhysPort, since nearly all funding has come through NSF lines targeting undergraduate education. According to the most recent studies conducted by AIP Statistics, there were 9,400 full-time equivalent (FTE) faculty members (including those in temporary or non-tenure-track positions) at physics degree-granting departments during the 2009-10 academic year[18] and "almost 3,300 faculty members taught physics courses in two-year colleges during the 2011-12 academic year"[19], giving about 13,000 U.S. college faculty in total. PhysPort has 2,600 Verified Educators who are U.S. college faculty, and 85% of those who give their department name are in a physics-related department; thus, it is estimated that about 20% of U.S. college physics faculty are Verified Educators. This suggests PhysPort is doing well at reaching its main target audience.

*2. U.S. high school physics teachers*

While PhysPort does not explicitly target high school physics teachers, many PhysPort resources are applicable to (or even designed for) high school classrooms. Since many AAPT members are high school teachers, advertising PhysPort via AAPT reaches many high school teachers. According to AIP Statistics, approximately 27,000 teachers in U.S. high schools taught at least one physics class during the 2012-13 school year[20]. PhysPort has 1,800 Verified Educators who are U.S. high school physics teachers, thus we estimate about 7% of U.S. high school physics teachers are Verified Educators. A logical future step for PhysPort would be to target high school teachers explicitly by conducting user research with them and creating resources targeted to the needs of high school classrooms.

*3. AAPT members*

AAPT is the main U.S. professional society for physics educators at all levels; as such, its 7,000 members are a natural target audience for PhysPort. PhysPort is regularly marketed to AAPT members through email announcements and presentations and workshops at meetings. AAPT members have the exclusive benefit of becoming Verified Educators instantly, while other users must wait 1-2 days for someone to verify their account. About 1,300 Verified Educators have used this "instant verification" process; therefore, we estimate that 20% of AAPT members are Verified Educators.[21] This is lower than we would like, so we see opportunities for further marketing to AAPT members. Additionally, because 80% of PhysPort Verified Educators are not AAPT members, there is also an opportunity for PhysPort to recruit new members to AAPT.

## What is the international impact of PhysPort?

PhysPort has many users outside the U.S: 31% of Verified Educators come from outside the U.S. Out of a total of 56,000 visitors to PhysPort in 2018, 40% were from outside the U.S (and 60% were US-based). It is beyond the scope of this paper to estimate the number of potential users outside the U.S. since data would need to be gathered from many different sources, but we expect that this number is at least an order of magnitude larger than the number of U.S. college faculty and high school teachers.

It is important to note that PhysPort was not designed for educators outside the U.S., who often teach in very different educational environments and cultures and speak different languages than in the U.S., and therefore may need different resources. Nevertheless, the PhysPort team is very interested in serving educators outside the U.S. and has taken preliminary steps in this direction. Our team has done some outreach and created several resources for users outside the U.S., including talks, workshops, and collaborations in Canada, Germany, Ghana, Kyrgyzstan, Mexico, Rwanda, and Tajikistan. Although primarily an English-language site, PhysPort includes some resources in other languages as well. In spite of very little effort to solicit translations or advertise the availability of translations on PhysPort, we receive offers each month to translate assessments. Twenty-seven of our 82 assessments are available in at least one language additional to English[22]; these represent 14% of all assessment downloads. The most translated assessment is the FCI, available in 29 languages additional to English. PhysPort hosts 12 assessments in Spanish, and we have an Expert Recommendation listing Spanish-language physics education resources[23].

Educators outside the U.S. are a large and important potential audience, but also a challenge to serve because their needs and contexts are so diverse, and because there is little U.S.-based funding available for international physics education work. We are very interested in future collaborations with physics educators abroad to expand the use and usefulness of PhysPort beyond the U.S.

## *Conclusions*

In the last decade, PhysPort has gone from an idea to a site recognized throughout physics education with over 80,000 visits and over 50,000 unique visitors in 2018, and 6,600 Verified Educators as of March 2019. PhysPort usage has been growing every year by key common measures. PhysPort has substantial usage among U.S. college faculty (its primary target audience), as well as U.S. high school teachers and educators outside the U.S. The Assessments and Expert Recommendations are the most popular sections of the site. Many users come back to PhysPort regularly and spend substantial time there, indicating that PhysPort is making a valued contribution to the field.

## Acknowledgments

PhysPort is wholly grant-supported, and has been created through multiple NSF grants: 0840853, 1256352/1256354, 1245490, 1347821/1347728, 1726113/1726479, and portions of 1323699, 1223405, 1323129, 1626496, 1140860, and 1140706. We thank the team that has made PhysPort possible.

| Site Section (release date) | Description |
|---|---|
| Assessments (2014) www.physport.org/assessments | Guides to over 90 research-based assessments in physics. Verified Educators can download assessments, answer keys, and implementation guides. |
| Expert Recommendations (2016) www.physport.org/recommendations | Short articles answering the most common questions physics instructors have about teaching and assessment. |
| Teaching Methods (2011) www.physport.org/methods | Guides to over 50 research-based teaching methods in physics. We define teaching methods broadly to include strategies, curricula, tools, and structures for courses. |
| Data Explorer (2016) www.physport.org/dataexplorer | Instant analysis and visualization of students' responses to multiple-choice research-based assessments in physics, and comparison to U.S. national averages. |
| Periscope (2015) www.physport.org/periscope | Video workshops for faculty professional development and training of Teaching Assistants and Learning Assistants, featuring video of student interactions in real classrooms and discussion questions for instructors. |
| Curricula (2013) www.physport.org/curricula | A collection of open-source research-based curricula hosted directly on PhysPort. |
| Virtual New Faculty Workshop (2015) www.physport.org/nfw | Video presentations from the popular New Faculty Workshop, where leaders in physics education research discuss their teaching methods and approaches. |
| Podcasts (2011) www.physport.org/podcasts | A series of podcasts about how instructors can use the results of PER in their classrooms. |

Table I. PhysPort content, in order from most frequently accessed.

| Target population (in order of priority[a]) | PhysPort Verified Educators from this population[b, c] | Total target population[d] | Percentage of target population using PhysPort (as Verified Educators) |
|---|---|---|---|
| U.S. College Physics Instructors | 2,600 | 13,000* | 20% |
| AAPT Members | 1,300 | 7,000 | 20% |
| U.S. High School Physics Teachers | 1,800 | 27,000* | 7% |
| Educators Outside the U.S. | 2,000 | Unknown | Unknown |

Table II. Target populations for PhysPort and percentage of each using PhysPort.
[a] The rationale for these priorities is explained in the main text.
[b] Verified Educators as of March 2019.
[c] Note that these populations overlap: AAPT members are typically also members of the other groups listed here.
[d] Total target population from AIP surveys (refs. 17, 18, 19); see main text.